# Glass-forming photoactive cholesteric materials doped by quantum dots: phototunable circularly-polarized emission.

Alexey Bobrovsky[1]*, Konstantin Mochalov[2], Vladimir Oleinikov[2], Valery Shibaev[1].

[1] Faculty of Chemistry, Moscow State University, Leninskie gory, Moscow, 119992 Russia, e-mail: bbrvsky@yahoo.com

[2] Shemyakin & Ovchinnikov Institute of Bioorganic Chemistry, Russian Academy of Sciences, 117871 Moscow, Russia

Cholesteric materials doped by small amounts of fluorescent compounds attract a great attention of researchers. Such interest has grown up because these materials have the unique emission properties.[1, 2] Cholesteric helical structure can be considered as one-dimensional photonic crystal having band gap with spectral position ($\lambda_{max}$) determined by value of helix pitch (P) and proportional to the average refractive index (n) (1):

$$\lambda_{max}=nP \qquad (1)$$

Pitch of the helix can be controlled by the chemical structure of substances used for materials preparation and by different external fields.[1, 2] One of the most promising tools for the controlling of helix pitch is the light action. In a number of papers the approaches for the creation of cholesteric low-molar-mass materials with photocontrollable helix pitch and band gap position were demonstrated.[3] Most of these approaches are based on introduction in the cholesteric materials chiral-photochromic fragments which are capable for photoisomerizing under light irradiation and drastic change of their helical twisting power. As a result of irradiation helix untwisting or twisting and shift of photonic band gap position take place.

In our paper [4] exploring this principle we have been, for the first time, prepared cholesteric materials on the base of cyclic siloxanes and sorbide chiral dopant with phototunable circularly polarized emission. UV irradiation and shift of selective light reflection peak position result in change of fluorescent dye emission intensity and degree of circular polarization characterized by dissymmetry factor $g_e$ (2):

$$g_e = 2(I_L - I_R)/(I_L + I_R) \qquad (2)$$

where $I_L$ and $I_R$ are the intensities of left- and right-handed circularly polarized light, respectively.

The similar approach has been used later for the creation of cholesteric materials possessing phototunable lasing properties.[5]

In all papers devoted to cholesteric fluorescent systems as fluorescent dopants the organic dyes were used. One of the strong disadvantages of these dyes is their instability caused by photobleaching of photodegradation. In order to avoid this disadvantage in the present paper we are for the first time used as fluorescent dopant quantum dots (QDs) or inorganic semiconductor particles having extremely high photostability, broad absorbance band, high quantum yields of emission and other promising characteristics.[6]

Despite the relatively large number of publications describing LC composites doped by different nanoparticles, [7] for our knowledge, there is only one recent paper [8] devoted to the study of circularly polarized light emission of QDs dispersed in cholesteric material. Taking into the account this fact, this our preliminary communication is devoted to the demonstration of phototunable circularly polarized fluorescent properties in cholesteric materials containing QDs as a fluorescent dopant.

As cholesteric matrix material we have chosen a glass forming cholesteric cyclosiloxane manufactured by Wacker Company. It displays a selective light reflection of left-handed circularly polarized light in the blue region of the spectrum ($\lambda_{max}$~ 450 nm). Clearing temperature of cyclosiloxane is 180-182 $^{o}$C; glass transition temperature ~50 $^{o}$C. As chiral-photochromic dopant responsible for helix photoshifting properties we have used 3.2 wt.% of derivative of isosorbide and cinnamic acid **Sorb**.[4, 9]

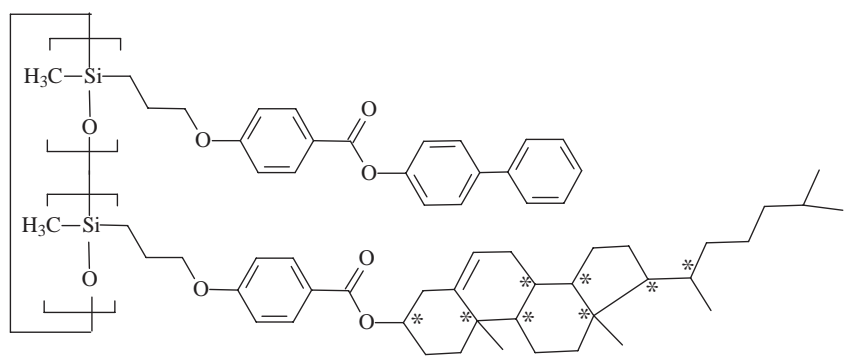

**SilBlue, 96.3%**

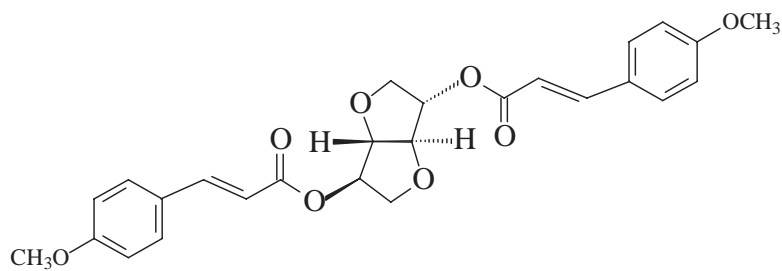

**Sorb, 3.2%**

This substance possesses a high helical twisting power and induces the formation of a right-handed cholesteric helical structure. UV-irradiation induces thermally irreversible E-Z isomerization of **Sorb** in respect of C=C bonds that is accompanied by a decrease of molecules anisometry (Fig. 1a) and lowering of helical twisting power. Introduction of this dopant into the cyclosiloxane matrix leads to a partial helix untwisting and to a shift of selective light reflection to the red spectral region (Fig. 1b).

Small amount of CdSe ZnS-coated quantum dots (0.5%) was added to the mixture. This type of inorganic nanoparticles has good fluorescent properties and is widely used as a component for nanocomposites preparation.[7]

The main goal of this communication is the description of the first results of optical and fluorescent properties study of QDs-containing cholesteric materials and the demonstration of possibility of photovariation of circularly polarized fluorescent properties in such materials.

Fig. 1b demonstrates the logarithm transmittance spectra of planarly-oriented samples of the initial mixture UV-irradiated during different time period. After each irradiation cycle the films were annealed at 130 $^o$C in order to achieve an equilibrium value of helix pitch (ca. 5 min) and selective light reflection wavelength; then they were slowly cooled down to room temperature (1 K/min) for excluding any mechanical deformation of the helical structure. As clearly seen from Fig. 1b, an irradiation results to the shift of selective light reflection peak to the shorter wavelength that is related to the decrease in the helical twisting power of **Sorb**.

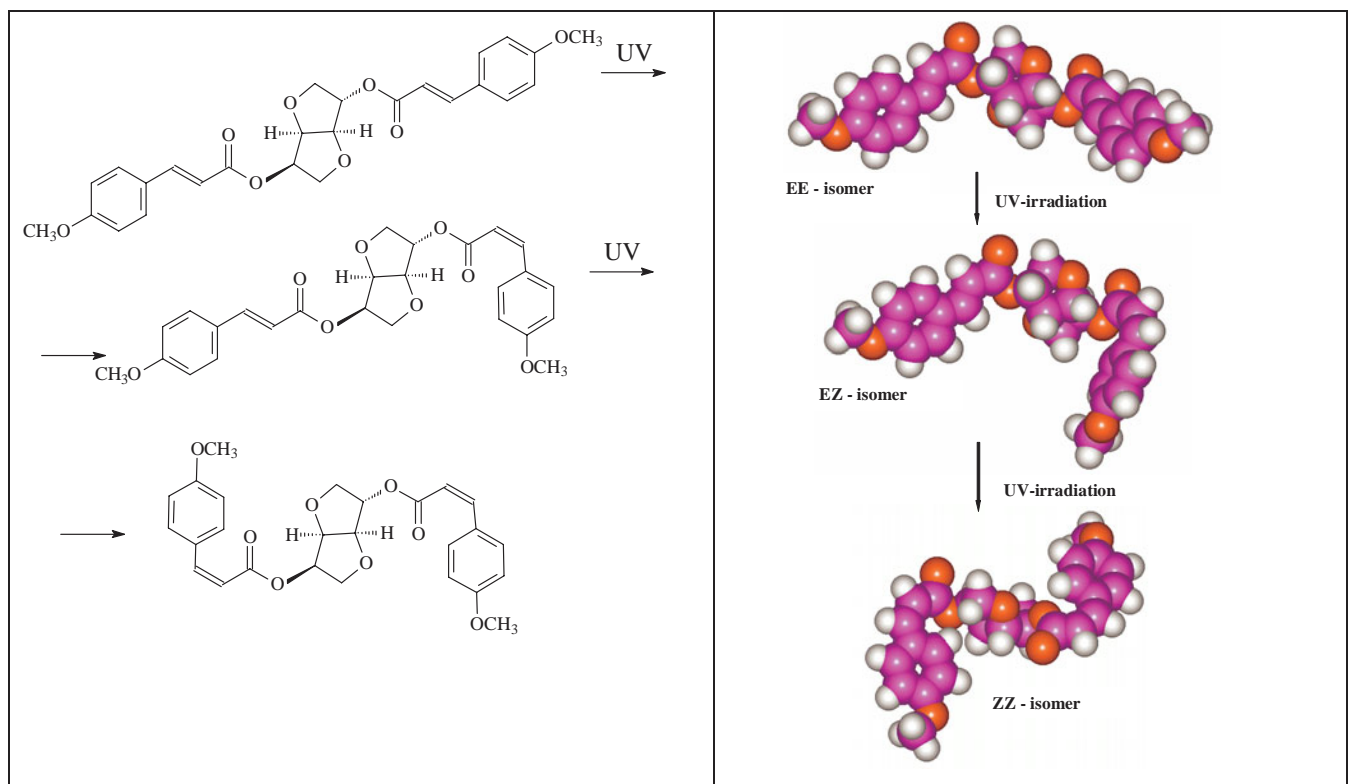

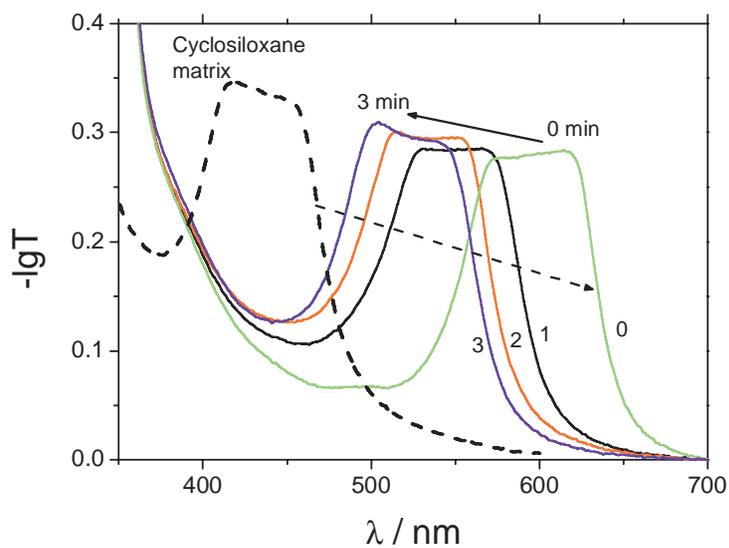

b

Fig. 1. (a) Change of **Sorb** molecules anisometry under UV-irradiation and E-Z isomerization; (b) logarithm transmittance spectra of irradiated samples (365 nm, ~2 mW/cm$^2$). The numbers of the curve correspond to the irradiation time.

QDs absorbance appears as a typical broad shoulder at wavelengths below ca. 500 nm which is clearly seen for films of mixture before UV-irradiation (Fig. 1b) and coincides with selective reflection band after helix twisting.

Let us consider how the shift of selective light reflection influences on the fluorescent properties of planarly oriented mixture films. Figs. 2a-d show spectra of left-handed circularly polarized component of emission of the nonirradiated and irradiated films together with corresponding logarithm transmittance spectra. Position of emission maximum is completely different for nonirradiated sample (518 nm) and film irradiated during 3 min (551 nm) (Figs. 3 a, d). In other words, UV-irradiation allows one to manipulate the fluorescence peak shape and emission colour. It is clearly seen, especially from Figs. 2b and c, that selective light reflection peak is completely coincided with intensity gap observed in emission spectra. It is noteworthy, that the right-handed circularly polarized component of fluorescence is not influenced by spectral position of the selective light reflection band. Both, position of the maximum as well as shape of the spectra in films are different from QDs solution in chlorophorm (dashed line in Fig. 1).

An appearance of two circularly polarized fluorescence peaks in Fig. 2b, c is related to the strong coinciding of photonic band with maximum of non-polarized fluorescence peak. That is why in this spectral region intensity of left-handed circularly polarized light emission decreases almost to zero, whereas two emission peaks are observed at the "shoulders" of photonic band gap.

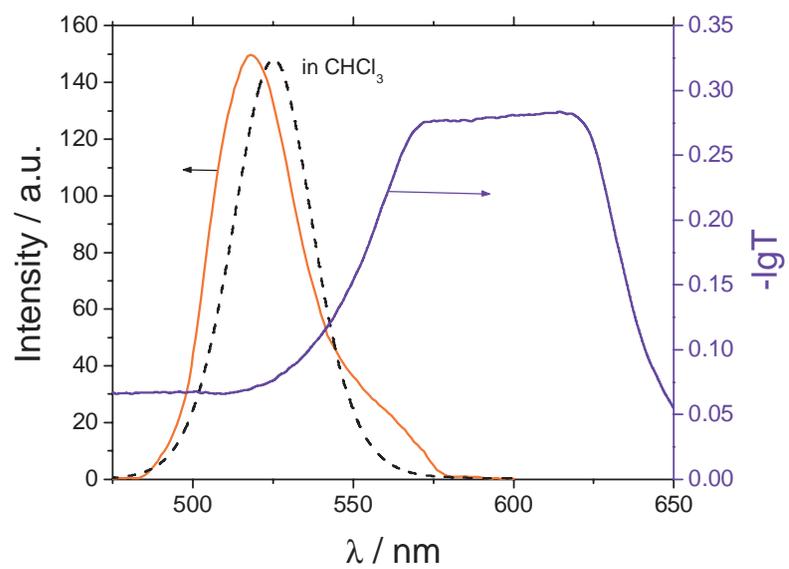

a

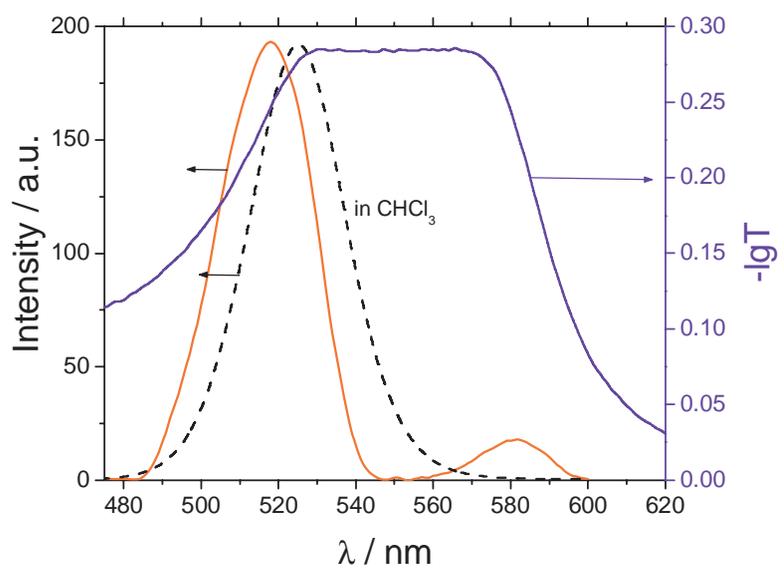

b

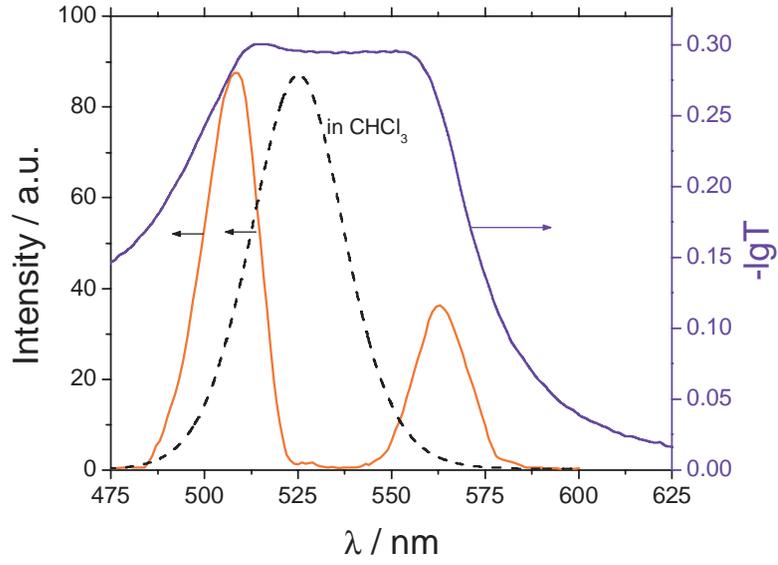

c

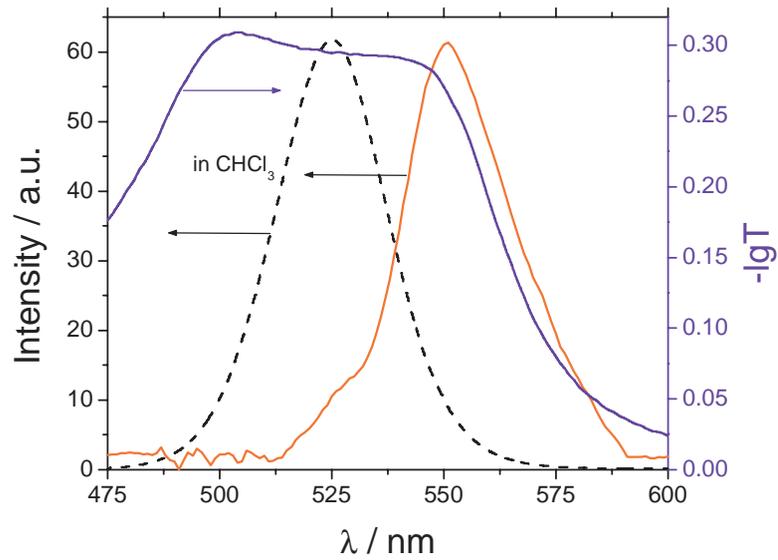

d

Fig. 2. Logarithm transmission and left-handed circularly polarized fluorescence spectra before (a) and after 1 (b), 2 (c), and 3 min (d) of UV irradiation followed by annealing at 130 °C (150 °C for 3 min). Excitation light wavelength is 400 nm. Dashed line shows normalized fluorescence spectra of QDs in chlorophorm solution.

Using the right- and left-handed circularly polarized fluorescence spectra and eq. (2) dissymmetry factor $g_e$ was calculated for films before and after UV-irradiation (Fig. 3). First of all, it is noteworthy that the values of dissymmetry factor are negative in the spectral regions corresponding to photonic band and completely coincides with the peaks of the selective light reflection and gradually shifts to the short-wavelength spectral region under UV-irradiation. Values of $g_e$ is very high and achieve the values of theoretical ones (-2) (see eq. 2) that demonstrates extremely high degree of circularly polarization of emitted light within selective light reflection peaks. Similar high values of dissymmetry factor were observed only in few papers [2] and indicate a good planar alignment of cholesteric materials in cells.

Secondly, at both shoulders of photonic bands there is a change of the sign of the dissymmetry factor. This effect concerns to a slight amplification of the left-handed emission components in these spectral ranges. It is interesting to note that for most organic fluorescent dyes such phenomenon is observed at long-wavelength shoulder of the photonic bands. It is explained by orientation of dyes molecules and their transition electronic moments along the director of liquid crystal matrix.[10] Thus, we may assume that in our case electronic transition moments of QDs are oriented randomly that promotes emission amplification at both shoulders.

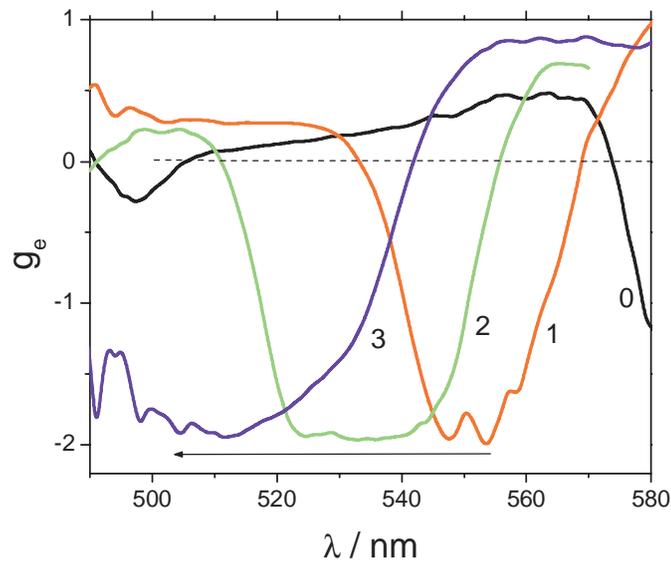

Fig. 3. Dissymmetry factor before, and after 1, 2 and 3 min of UV irradiation followed by annealing at 130 °C. Shift direction of the peaks is shown by arrow; the numbers of the curve correspond to the irradiation time in min.

In summary, for the first time, we have obtained cholesteric materials with photochemically tunable intensity and degree of circular polarization emission of quantum dots embedded in glass-forming matrix. Our future effort will be focused on the study of relations between size and shape of nanoparticles and photooptical properties of such systems and investigations of lasing phenomena.

*Acknowledgements.*

This research was supported by the Russian Foundation of Fundamental Research (11-03-01046, 09-03-12234-ofi-m) and Program COST-D35.

**Experimental Section.**

*Synthesis of CdSe quantium dots*

QDs of the core/shell structure (CdSe core and ZnS shell) were synthesized according to the standard method.[11] Briefly, an appropriate amount of trioctylphosphine oxide (Aldrich Chemical, Milwaukee, Wisconsin) was heated to 180° C under argon, dried, and degassed at this temperature under vacuum. Then, trioctylphosphine oxide was heated to 340° C under argon flow and intensive stirring. Solutions of dimethylcadmium (Strem, Bischheim, France) and selenium (elemental powder; Aldrich) precursors in trioctylphosphine (Fluka, Buchs, Switzerland) were injected through syringe in less than 1 second. The reaction was cooled to 300° C and a 1 M solution of dimethylzinc (Aldrich) in heptane and a solution of hexamethyldisilthiane (Fluka) were added dropwise under vigorous stirring. The reaction mixture was cooled to 50° C, and NCs were precipitated out from the solution by adding anhydrous methanol followed by centrifugation at 14,000 rpm. The precipitate was washed with methanol, and QDs were dissolved in chloroform (Sigma Chemical, St. Louis, Missouri). The procedure that we describe permits the synthesis of homogeneous QDs emitting the fluorescence from nearly 500 nm to 620 nm depending on their diameters. For this article we have used perfectly homogeneous QDs possessing a CdSe core of approximately 3.6 nm in diameter with an epitaxially grown ZnS shell of several monolayers in thickness.

Fluorescence spectrum of QDs used in this work is shown in Fig. 1a ($\lambda_{max}$=525 nm).

The phase transition temperatures of the mixture were determined by the polarizing optical microscope investigations were performed using LOMO P-112 polarizing microscope equipped by Mettler TA-400 heating stage. Clearing temperature of the obtained mixture is 171-175 $^{\circ}$C that is about 10 degrees lower than for the pure cyclosiloxane used as a matrix; glass transition temperature is the same as for cyclosiloxane (~50 $^{\circ}$C). An aggregation and phase separation does not take place during the mixture preparation and mixture is quite homogeneous. Absence of nanoparticles aggregation was confirmed by results of fluorescent microscopy. It is noteworthy that doping of cyclosiloxane by chiral-photochromic dopant and quantum dots does not lead to any noticeable changes in optical quality of planarly-oriented films which remain rather transparent.

The optical and photooptical studies 25-µm-thick films sandwiched between two flat glass plates were prepared. The thickness of the test samples was controlled by glass beads. For preparation of good planar texture we have used a LC photoalignment technique. Glass plates were spin-coated by solution of poly[1-[4-(3-carboxy-4-hydroxy-phenylazo) benzenesulfonamido]-1,2-ethanediyl, sodium salt (Aldrich), 2 mg/mL. After drying at room temperature glass plates coated by this polymer were irradiated by polarized polychromatic light of mercury lamp (~15 mW/cm$^2$, 30 min). Glan-Taylor prism was used as polarizer. A planar texture was obtained by shear deformation of the samples, which were heated up to temperatures well above the glass transition temperature (130 $^{\circ}$C). After irradiation prior to performing absorbance and fluorescence measurement, the samples were annealed for about 20 min followed by cooling down to room temperature at the rate 1 $^{\circ}$/min.

Transmittance spectra of planarly-oriented films were recorded by Unicam UV-500 UV-Vis spectrophotometer.

Photochemical investigations were performed using an optical set up equipped with a DRSh-250 ultra-high pressure mercury lamp. To prevent the heating of the samples due to the IR irradiation of the mercury lamp, a water filter was introduced in the optical scheme. To assure the plane-parallel light beam, a quartz lens was applied. Using the filter a light with the wavelengths 365 nm were selected. The intensity of light was measured by LaserMate-Q (Coherent) intensity meter (~2 mW/cm$^2$).

Fluorescence spectra were recorded using Shimadzu RF-5301PC spectrofluorophotometer with the detection normal to the plane of the film whereas the excitation beam was positioned at the certain angle from the back side of the films. Circularly-polarized absorbance and fluorescence spectra were obtained by using a combination of a linear polarizer with a broad-band quarter-wave plate.

**References**

[1] B.M. Conger, J.C. Mastrangelo, S.H. Chen, *Macromolecules* **1997**, *30*, 4049; B.M. Conger, D. Katsis, J.C. Mastrangelo, S.H. Chen, *J. Phys. Chem. A* **1998**, *102*, 9213; H. Shi, B.M. Conger, D. Katsis, S.H. Chen, *Liq. Cryst*. **1998**, *24*, 163; S.H. Chen, D. Katsis, A.W. Schmid, J.C. Mastrangelo, T. Tsutsuik, T.N. Blanton, *Nature* **1999**, *397*, 506; D. Katsis, D.U. Kim, H.P. Chen, L.J. Rothberg, S.H. Chen, *Chem. Mater*. **2001**, *13*, 643; M. Voigt, M. Chambers, M. Grell, *Liq. Cryst*. **2002**, *29*, 653; R.K. Vijayaraghavan, S. Abraham, H. Akiyama, S. Furumi, N. Tamaoki, S. Das, *Adv. Funct. Mater*. **2008**, *18*, 1.

[2] S.M. Jeong, Y. Ohtsuka, N.Y. Ha, Y. Takanishi, K. Ishikawa, H. Takezoe, S. Nishimura, G. Suzaki, *Appl. Phys. Lett*. **2007**, *90*, 211106; K.L. Woon, M. O'Neil, G.J. Richards, M.P. Aldred, S.M. Kelly, A.M. Fox, *Adv. Mater*. **2003**, *15*, 1555.

[3] E. Sackmann, *J. Am. Chem. Soc*. **1971**, *93*, 7088; Feringa, B. L.; van Delden, R. A.; Koumura, N. Geertsema, E. M. *Chem. Rev*. **2000,** *100,* 1789; Shibaev, V.; Bobrovsky, A.; Boiko, N. *Prog. Polym. Sci*. **2003**, *28*, 729; Tamaoki, N. *Adv. Mater*. **2001**, *13*, 1135; Van de Witte, P.; Galan, J. C.; Lub, J., *Liq. Cryst*. **1998**, *24*, 819; Bobrovsky, A. Yu.; Boiko, N. I.; Shibaev, V. P. *Liq. Cryst*. **1998**, *25*, 679.

[4] A. Bobrovsky, N. Boiko, V. Shibaev, J. Wendorff, *Adv. Mater.* **2003**, *15,* 282.

[5] P.V. Shibaev, R.L. Sanford, D. Chiappetta, V. Milner, A. Genack, A. Bobrovsky, *Opt. Expr.* **2005**, *13*, 2358; G. Chilaya, A. Chanishvili, G. Petriashvili, R. Barberi, R. Bartolino, G. Cipparrone, A. Mazzulla, P.V. Shibaev, *Adv. Mater.* **2007**, *19*, 565.

[6] B.O. Dabbousi, J. Rodriguez-Viejo, F.V. Mikulec, J.R. Heine, H. Mattoussi, R. Ober, K.F. Jensen, M. G. Bawendi, *J. Phys. Chem. B* **1997**, *101,* 9463; A.P. Alivisatos, *Science* **1996**, *271*, 933.

[7] L.S. Hirst, J. Kirchhoff, R. Inmana, S. Ghosh, *Proc. of SPIE* **2010**, 76180F; L.C.T. Shoute, D.F. Kelley, *J. Phys. Chem. C* **2007**, *111,* 10233; J.-F. Blach, S. Saitzek, C. Legrand, L. Dupont, J.-F. Henninot, M. Warenghem, *J. Appl. Phys. L*ett. **2010**, *107*, 074102; V. Domenici, B. Zupancic, V.V. Laguta, A.G.

Belous, O.I. Vyunov, M. Remskar, B. Zalar, *J. Phys. Chem. C* **2010**, *114,* 10782; L. Cseh, G.H. Mehl, *J. Mater. Chem.* **2007**, *17*, 311; S. Khatua, P. Manna, W.-S. Chang, A. Tcherniak, E. Friedlander, E.R. Zubarev, S. Link, *J. Phys. Chem. C* **2010,** *114,* 7251; R. Pratibha, W. Park, I.I. Smalyukh, *J. Appl. Phys.* **2010**, *107*, 063511.

[8] S.G. Lukishova, L.J. Bissell, C.R. Stroud, Jr., R.W. Boyd, *Optics and Spectroscopy,* **2010***, 108,* 417.

[9] A.Yu. Bobrovsky, N.I. Boiko, V.P. Shibaev, *Mol. Cryst. Liq. Cryst.,* **2001**, *363*, 35; A.Yu. Bobrovsky, N.I. Boiko, V. P. Shibaev, M.A. Kalik, M.M. Krayushkin, *Pol. Adv. Techn.,* **2002**, *13*, 595.

[10] J. Schmidtke, W. Stille, *Eur. Phys. J. B* **2003**, *31*, 179.

[11] M.A. Hines, P. Guyot-Sionnest, *J. Physiol. Chem.* **1996**, *100*, 468.